\documentclass[12pt]{article}
\usepackage{amssymb,amsmath,epsfig}
\allowdisplaybreaks
\begin{document}

\title{\bf Evolution of Cluster of Stars in $f(R)$ Gravity}

\author{Rubab Manzoor $^1$
\thanks{rubab.manzoor@umt.edu.pk;~~dr.rubab.second@gmail.com}
and Wasee Shahid $^2$
\thanks{wasee.mathematician@gmail.com}\\
Department of Mathematics,\\ University of Management and
Technology,\\
Johar Town Campus,\\ Lahore-54782, Pakistan.}
\date{}
\maketitle

\begin{abstract}
This paper explores the dynamics of evolving cluster of stars in the
presences of exotic matter. The $f(R)$ theory is used to presume
exotic terms for evolution scenario. We use structure scalars as
evolution parameters to explore dynamics of spherically symmetric
distribution of evolving cluster of stars. We consider Starobinsky
model, $f(R)=R+\epsilon R^2$ and study different evolution modes
having features like isotropic pressure, density homogeneity,
homologous and geodesic behavior. It is concluded that dynamics of
these modes of evolution depends upon the behavior of dark matter.
The presences of dark matter directly affects the features of
cluster like anisotropic pressure, dissipation, expansion,
shear as well as density homogeneity.
The evolution having homogeneous density and
isotropic pressure depends upon conformally flat and non-dissipative
behavior of baryonic as well as non baryonic matter. The dissipation
factor induces density inhomogeneity in the expanding clusters having
shear effects.  The non dissipative homologous evolution also be
discussed in the presence and absence of shear effects. It is found
that high curvature geometry in the presence of shear supports
homologous evolution. Expanding clusters
are also explored in the presences of dissipation of dark matter
and shear effects. By using quasi-homologous conditions the
geodesic evolution is studied. It is theoretically showed that
geodesic and homologous conditions depends upon each other.
\textbf{Finally we investigate behavior of Starobinsky model for a stellar
structure $4U 1820-30$ toward center.
It is found that for increasing values of $\epsilon$ the DM
behavior is dominant as compare to baryonic matter.
}\end{abstract}

\section{Introduction}

The mystery of dark energy and dark matter is one of the important
topic of modern physics. Recently, Planck's \cite{1a} gives
distribution which shows $5\%$ baryonic matter with $27\%$ dark
matter (DM) and $71\%$ dark energy (DE). It is believed that DE is
responsible for the expansion of the universe in accelerated rate as
time is passing. The matter that can be seen easily is known as
baryonic matter. While the DM or non-baryonic matter is an invisible
hypothetical type of matter which does not or very weakly interact
with electromagnetic radiations. The DM is observable by its
gravitational effects on baryonic matter. Since the rise of last
century, a sensational amount of evidence indicating a dominant
contribution of DM to the universe. Different detection mechanism
have been constructed to study the nature of DM. Among these, the
search for appearing of DM in stellar bodies comes as inexpensive
and ingenious detection strategy. Naturally, in this context, most
indirect searches have been focused yielding fruitful results that
are complementary and competitive.

In order to explore mystery of dark universe (DE and DM) many
candidates are introduced by researchers. In this context, the
$f(R)$ theory of gravity \cite{4a} is one of the most explored
example involving higher order curvature terms. This theory is the
generalization of Einstein Hilbert action with a generic function of
curvature ($f(R)$). The $f(R)$ gravity is very interesting because
of its higher order curvature terms, which can represents both DE
and DM problems sufficiently. Many researchers \cite{3ab,4ab} used
various $f(R)$ models to study the effects of DE and DM. The
Starobinsky model ($f(R)=R+\epsilon R^2$) \cite{12a,8ab} is a model
of gravity in power-law type of the Ricci scalar that has
efficiently promoted the early-time cosmic expansion and DM issues.

A stars cluster is a group of stars which are gravitationally bound
together. The average stars distance in a cluster is much closer
than average distance between stance and rest of the galaxy. It is
believed that study of stars cluster can be used to examine the
evolution and formation of galaxies. This in turn describes
evolution and formation of the galactic system. There have been a
meaningful theoretical efforts to study the evolution and formation
of stars cluster \cite{6ab}. In this context, it would be
interesting to discuss the evolution of stars cluster through the
technique of structure scalars.

Structure scalars are scalar expressions which deals with mixture of
physical terms like density, anisotropy of pressure as well as Weyl
tensor. These scalars are used to describe the evolution of
self-gravitating systems. Firstly, Herrera et al. \cite{3i}
suggested the orthogonal division of the Riemann curvature tensor
and found several scalars of structure. The basic properties of
stellar are then correlated with these scalars. By using structure
scalars, they interpret some properties of relativistic structures.
They studied hydrodynamics along with thermodynamics of
self-gravitating systems like Lemaitre-Tolman-Bondi space time
\cite{4ia}, spherical model with electric charge and cosmological
constant \cite{4i}, cylindrically as well as axially symmetric
self-gravitating systems \cite{7ab}. Herrera et al. \cite{6abb}
discussed evolution of complex less stellar systems based on
baryonic matter with the help of structure scalars. In modified
theories, Sharif and Manzoor \cite{14a} and Yousaf and his
collaborators \cite{6i} applied these findings to different cosmic
models.

The DM is an important ingredient of the cluster of stars \cite{2a}.
It is well known about $95\%$ of the total mass of the stars cluster
is made up of DM. The problems of galactic clusters, galactic
rotational curves and the mass discrepancy indicate the importance
and existence of DM in the evolution of stars cluster \cite{2aa}. It
is believed that the presence of DM affects the stellar evolution.
So the study of evolving stellar like clusters in the presence of DM
may determinate the dynamics of its evolving process. In this way,
we may get about modifications that are not visible in the structure
of the universe \cite{14a,3a}.  In the current study, we discuss
evolution of cluster of stars in the presence of DM. In this regard,
we use $f(R)$ gravity as DM candidate and use the technique of
structure scalars to explore different models of evolution of
clusters of stars. The paper is organized as: the next section
defines the $f(R)$ gravity. Section \textbf{3} describes spherically
symmetric cluster of stars. Section \textbf{4} shows structure
scalars as a evolution parameters. Section \textbf{5} applies
Starobinsky model as a DM candidate to the cluster framework.
Section \textbf{6} discusses various types of cluster evolutions.
Finally, section \textbf{7} summarizes the results.

\section{$f(R)$ formalism}

The Einstein-Hilbert action is
\begin{equation}\label{1aB}
S_{EH}=\frac{1}{2k}\int d^4x\sqrt{-g}R,
\end{equation}

where $k\equiv8\pi G,~ G$ is the gravitational constant and $g$ is
determinant of the metric tensor and $R$ is the Ricci scalar. The
action of $f(R)$ gravity generalizes Einstein-Hilbert action as
\begin{equation}
S=\frac{1}{2k}\int d^4x\sqrt{-g}f(R),
\end{equation}
where $f(R)$ is a generic algebric function of the Ricci scalar. The
variation of the above action with respect to metric tensor provides
modified field equations
\begin{equation}\label{1B}
R_{\alpha\beta}f_{R}-\frac{1}{2}f(R)g_{\alpha\beta}+(g_{\alpha\beta}\Box
- \bigtriangledown_{\alpha}
\bigtriangledown_{\beta})f_{R}=kT_{\alpha\beta}.
\end{equation}
Here $\bigtriangledown_{\alpha}$ is the covariant derivative,
$\Box\equiv\bigtriangledown^{\beta}\bigtriangledown_{\beta}$ be the
D-Alembert operator and $f_R\equiv \frac{df}{dR}$. After some
manipulation the field equations turn out to be
\begin{equation}\label{1*B}
G_{\alpha\beta}=\frac{k}{f_{R}}\left(T_{\alpha\beta}+T_{\alpha\beta}^{D}\right),
\end{equation}
where
\begin{equation}\nonumber
T_{\alpha\beta}^{(D)}={\bigtriangledown_{\alpha}
    \bigtriangledown_{\beta}f_{R}+\left(f-Rf_{R}\right)
    \frac{g_{\alpha\beta}}{2}-\Box f_{R}g_{\alpha\beta}}.
\end{equation}

\section{General setup of Problem}

We assume an evolving cluster of stars in the form of spherically
symmetric fluid distribution where the fluid particles are
combination of baryonic matter (stars) and non-baryonic matter (DM).
The interior geometry of cluster is represented by a spherically
symmetric line element
\begin{equation}\label{2B}
ds^2=-A^2(t,r) dt^2 +B^2(t,r) dr^2 +C^2(t,r) d\theta^2 +
C^2(t,r)sin^2 \theta d\phi^2.
\end{equation}
Here $A,~B,~C$ are considered positive with $A,~B$ as dimensionless
terms whereas $C$ and $r$ have same dimensions.  We specify the
coordinates $x^0,x^1, x^2$ and $x^3$ as $t,~r,~\theta$ and $\phi$,
respectively.

We consider the fluid distribution to be anisotropic and
dissipative, demonstrated by the energy-momentum
$T^{(eff)}_{\alpha\beta}$, given by
\begin{eqnarray}\nonumber
T^{(eff)}_{\alpha\beta}&=&(\mu ^{(eff)}+P_\perp^{(eff)}) V_\alpha
V_\beta+P_\perp^{(eff)}g_{\alpha\beta}+(P_{r}^{(eff)}-P_\perp^{(eff)})
\chi_\alpha \chi_{\beta}\\\label{3B}
&+&q_{\alpha}^{(eff)}V_{\beta}+V_{\alpha} q_{\beta}^{(eff)}.
\end{eqnarray}
Here $P^{(eff)}_r$ shows radial pressure, $P_\perp^{(eff)}$ is
tangential pressure, $\mu^{eff}$ indicates effective energy density
and $q_\alpha^{(eff)}$ describes heat flux. The term $V^\alpha$
represents four velocity of the fluid, $\chi^\alpha$ shows unit four
vector towards radial direction define as (for a comoving observers)
\begin{equation}\label{aB*}
V^\alpha=A^{-1}\delta_0^\alpha,\quad
q^{(eff)\alpha}=q^{(eff)}B^{-1}\delta^\alpha_1,\quad
\chi^\alpha=B^{-1}\delta^\alpha_1
\end{equation}
and satisfies
\begin{equation}
V^\alpha V_\alpha=-1,~V^\alpha q_\alpha=0,~\chi^\alpha
\chi_\alpha=1,~ \chi^\alpha V_\alpha=0.
\end{equation}

The canonical form of energy momentum tensor appropriated as
\begin{equation}\label{4B}
T_{\alpha\beta}^{(eff)}=\mu^{(eff)}  V_\alpha V_\beta +
Ph_{\alpha\beta}^{(eff)}+\Pi^{(eff)}_{\alpha\beta}+q^{(eff)}
(V_\alpha \chi_\beta +\chi_\alpha V_\beta),
\end{equation}
where
\begin{eqnarray}\nonumber
P&=&\frac{P_r^{(eff)}+2P_{\perp}^{(eff)}}{3},\quad
h_{\alpha\beta}=g_{\alpha\beta}+V_{\alpha V_\beta},\\\nonumber
\Pi^{(eff)}_{\alpha\beta}&=&\Pi^{(eff)}(\chi_\alpha
\chi_\beta-\frac{1}{3}h_{\alpha\beta}),\quad
\Pi^{(eff)}=P_r^{(eff)}-P^{(eff)}_{\perp}.
\end{eqnarray}
The modified field equations becomes
\begin{eqnarray}
&G_{00}=\mu^{(eff)}=\frac{k}{f_{R}}\left[\mu^{m} A^{2}+\frac{A^{2}}{K}\left\{\frac{f-R f_{R}}{2}+\frac{f_{R}^{\prime \prime}}{B^{2}}+\left(\frac{2 \dot{C}}{C}-\frac{\dot{B}}{B}\right)\frac{\dot{f}_{R}}{A^{2}}\right.\right.\\\label{5B}
&\left.\left.+\left(\frac{2 C^{\prime}}{C}-\frac{B^{\prime}}{B}\right) \frac{f_{R}^{\prime}}{B^{2}}\right\}\right]\\\nonumber
&G_{11}=P_{r}^{(e f f)}=\frac{K}{f_{R}}\left[\operatorname{P^{m}_r} B^{2}-\frac{B^{2}}{K}\left\{\frac{f-R f_{R}}{2}-\frac{\ddot{f}_{R}}{A^{2}}+\left(\frac{2 \dot{C}}{C}+\frac{\dot{A}}{A}\right) \frac{\dot{f}_{R}}{A^{2}}\right.\right.\\\label{6B}
&\left.\left.+\left(\frac{2 C^{\prime}}{C}+\frac{A^{\prime}}{A}\right) \frac{f_{R}^{\prime}}{B^{2}}\right\}\right]\\\nonumber
&G_{22}=P_{\perp}^{(eff)}=\frac{k}{f_{R}}\left[P^{m}_{\perp} C^{2}-\frac{C^{2}}{K}\left\{\frac{f-R f_{R}}{2}+\frac{f_{R}^{\prime \prime}}{B^{2}}-\frac{\ddot{f}_{R}}{A^{2}}+\left(\frac{\dot{C}}{C}-\frac{\dot B}{B}+\right.\right.\right.\\\label{7B}
&\left.\left.\left.\frac{\dot{A}}{A}\right)
\frac{\dot{f}_{R}}{A^{2}}+\left(\frac{2 C^{\prime}}{C}+\frac{A}{A}\right)
\frac{f_{R}^{\prime}}{B^{2}}\right\}\right],\\\label{8*B}
&G_{01}=q^{(eff)}=\frac{K}{f_R}(T_{01}+T_{01}^D)=-\frac{1}{f(R)}(f_{R;r;t}
-\frac{2A^\prime}{A}\dot{f_R}-\frac{\dot{B}}{B}f_R^\prime
\\\nonumber
+KABq^{m}).
\end{eqnarray}
Here superscript $m$ shows energy term associated to baryonic matter
and prime as well as dot represent the differentiation with respect
to $r$ and $t$, respectively. It can be noticed that the shear
viscosity trivially absorb into the radial and tangential pressures,
so it cannot be added to the system explicitly. Similarly,
dissipation in the free streaming approximation cannot be expressed
individually, because it merges in $\mu^{(eff)}$, ~$P_r^{(eff)}$ and
$q^{(eff)}$.

The acceleration $a_\alpha$ and the expansion $\Theta$ of the fluid
are evaluated as
\begin{equation}\label{bB*}
a_\alpha=V_{\alpha;\beta}V^\beta,\quad \Theta=V^\alpha_{;\alpha}.
\end{equation}
The shear tensor $\sigma_{\alpha\beta}$ is
\begin{equation}\label{cB*}
\sigma_{\alpha\beta}=V_{(\alpha;\beta)}+a_{(\alpha}V_{\beta)}-\frac{1}{3}\Theta
h_{\alpha\beta}.
\end{equation}
By using (\ref{bB*}) with (\ref{aB*}), the obtain expressions of
non-vanishing component of four-acceleration and its scalar are
\begin{equation}\label{8B}
a_1=A^{-1}{A^\prime},\quad a=\sqrt{a^\alpha
    a_\alpha}=\frac{A^\prime}{AB},
\end{equation}
and the expansion scalar is calculated as
\begin{equation}\label{9B}
\Theta=A^{-1}\left(\frac{\dot{B}}{B}+2\frac{\dot{C}}{C}\right).
\end{equation}

Using Eqs.(\ref{aB*}) and (\ref{cB*}), we obtain the non-zero
components of shear tensor and shear scalar given by
\begin{equation}\label{10B}
\sigma_{11}=\frac{2}{3}B^2\sigma,\quad\sigma_{22}=
\frac{\sigma_{33}}{sin^2\theta}=-\frac{1}{3}C^2\sigma,\quad
\sigma^{\alpha\beta}\sigma_{\alpha\beta}=\frac{2}{3}\sigma^2,
\end{equation}
where
\begin{equation}\label{11B}
\sigma=\frac{1}{A}\left(\frac{\dot{B}}{B}-\frac{\dot{C}}{C}\right).
\end{equation}

The Misner Sharp mass function $m(t,r)$ describes the total energy
of the spherical system within radius ``$C$" and reads as
\begin{equation}\label{12B}
m=\frac{C^3}{2}R^{23}_{23}
=\frac{C}{2}\left[\left(\frac{\dot{C}}{A}\right)^2
-\left(\frac{C^\prime}{B}\right)^2+1\right].
\end{equation}
The proper time derivative $D_T$ is given by
\begin{equation}\label{13*B}
D_T=\frac{1}{A}\frac{\partial}{\partial t},
\end{equation}
and the variation of areal radius $C$ (inside the fluid) with
respect to proper time provide the velocity of collapsing fluid
given by
\begin{equation}\label{14B}
U=D_T C < 0.
\end{equation}
Now, equation (\ref{12B}) can be rewritten as
\begin{equation}\label{15B}
E\equiv \frac{C'}{B}=\left(1+U^2-\frac{2m}{C}\right)^\frac{1}{2}
\end{equation}
using (\ref{9B}), (\ref{11B}) and (\ref{15B}), Eq.(\ref{8*B}) can be
rewritten as
\begin{equation}\label{26*B}
4\pi
q^{(eff)}=E\left[\frac{1}{3}D_C(\Theta-\sigma)-\frac{\sigma}{C}\right],
\end{equation}
where $D_C$ denotes the proper radial derivative given by
\begin{equation}\label{27*B}
D_C=\frac{1}{C^\prime}\frac{\partial}{\partial r}.
\end{equation}
Using field equations (\ref{5B})-(\ref{8*B}), with (\ref{13*B}) and
(\ref{27*B}), we get from (\ref{12B})
\begin{equation}\label{28B}
D_T m=-4\pi \left(P_r^{(eff)}U+q^{(eff)}E\right)C^2,
\end{equation}
\begin{equation}\label{29B}
D_C m=4\pi\left(\mu^{(eff)}+q^{(eff)}\frac{U}{E}\right)C^2,
\end{equation}
which yield
\begin{equation}\label{30B}
m=4\pi \int^r_0 \left(\mu^{(eff)}+q^{(eff)}\frac{U}{E}\right)C^2 C'
dr.
\end{equation}
Integrating above equation along with condition $m(t,0)=0$ gives
\begin{equation}\label{31B}
\frac{3m}{C^3}=4\pi\mu^{(eff)}-\frac{4\pi}{C^3}\int^r_0
C^3\left(D_{C}\mu^{(eff)}-3q^{(eff)}\frac{U}{CE}\right)C'dr.
\end{equation}

The Weyl tensor $C_{\alpha\mu\beta\nu}$ describes the effect of
tidal forces and evaluated in terms of Riemann tensor, Ricci tensor
and Ricci scalar. For spherically symmetric distribution the
magnetic part of the Weyl tensor vanishes and the electric part of
Weyl tensor $E_{\alpha\beta}$ \cite{8b} is given by
\begin{equation}\label{33*B}
C_{\alpha\mu\beta\nu}=(g_{\alpha\mu\gamma\vartheta}g_{\beta\nu\delta\sigma}
-\eta_{\alpha\mu\gamma\vartheta}\eta_{\beta\nu\delta\sigma})V^{\gamma}V^{\delta}E^{\vartheta\sigma},
\quad
g_{\alpha\mu\gamma\vartheta}=g_{\alpha\gamma}g_{\mu\vartheta}-g_{\alpha\vartheta}g_{\gamma\mu},
\end{equation}
where $\eta_{\alpha\mu\gamma\vartheta}$ is Levi-Civita tensor.
Equation (\ref{33*B}) implies
\begin{equation}\label{32B}
E_{\alpha\beta}=C_{\alpha\mu\beta\nu}V^\mu V^\nu.
\end{equation}
The non-vanishing components of $E_{\alpha\beta}$ are
\begin{equation}\label{33B}
E_{11}=\frac{2}{3}B^2\varepsilon,\quad
E_{22}=-\frac{1}{3}R^2\varepsilon,\quad E_{33}=E_{22}Sin^2\theta
\end{equation}
with
\begin{eqnarray}\label{34*B}
\varepsilon &=&\frac{1}{2A^2}\left[\frac{\ddot
    {C}}{C}-\frac{\ddot{B}}{B}-\left(\frac{\dot{C}}{C}-
\frac{\dot{B}}{B}\right)\left(\frac{\dot{A}}{A}+\frac{\dot{C}}{C}\right)\right]+
\frac{1}{2B^2}\left[\frac{A''}{A}-
\frac{C''}{C}+\left(\frac{B'}{B}+\frac{C'}{C}\right)\right.\nonumber
\\ &&\left.\left(\frac{C'}{C}-\frac{A'}{A}\right)\right]-\frac{1}{2C^2}.
\end{eqnarray}
The electric part of Weyl tensor can be also expressed as
\begin{equation}\label{35B}
E_{\alpha\beta}=\varepsilon\left(\chi_\alpha \chi_\beta
-\frac{1}{3}h_{\alpha\beta}\right).
\end{equation}

\subsection{The Exterior Space time}

We assume a bounded cluster of stars. In this condition, the
junction condition have to satisfied at hypersurface $\Sigma$. Thus,
we consider Vaidya spacetime as exterior geometry at $\Sigma$ (i.e
radiations have no mass which are going outside)
\begin{equation}\label{49B}
ds^2=-\left[1-\frac{2M(v)}{r}\right]dv^2-2drdv+r^2(d\theta^2 +
sin^2\theta d\phi^2).
\end{equation}
By comparing the full non-adiabatic sphere to the Vaidya spacetime,
when $r=r_{\Sigma}$ becomes constant, we get
\begin{equation}\label{50B}
m(t,r)= M(v),
\end{equation}
and
\begin{eqnarray}\label{51B}
2\left(\frac{\dot C^\prime}{C}-\frac{\dot
    B}{B}\frac{C^\prime}{C}-\frac{\dot C}{C}\frac{A^\prime}{A}\right)&=&
- \frac{B}{A}\left[2\frac{\ddot C}{C}-\left(2\frac{\dot
    A}{A}-\frac{\dot C}{C}\right)\frac{\dot
    C}{C}\right]+\frac{A}{B}\left[\left(2\frac{A^\prime}{A}+\frac{C^\prime}{C}\right)\right.\nonumber
\\&&\left.\frac{C^\prime}{C}-\left(\frac{B}{C}\right)^2\right].
\end{eqnarray}

\section{The Structure Scalars}

The structure scalars are scalar functions that are derived through
the concept of orthogonal splitting of Riemann tensor \cite{3i} and
are used to explore dynamics of systems involving self-gravitating
bodies. These factors suggest refine mechanism in stellar evolution
by reducing complexity of the system \cite{6abb}. In order to
suggest the dynamics of refine (complex less) evolution of
collection of stars, we use the generalized concept of structure
scalars in $f(R)$ gravity.

According to orthogonal splitting of the Reimann tensor \cite{3i},
the tensors $Y_{\alpha\beta}$ and $X_{\alpha\beta}$ are given by
\begin{equation}\label{36B}
Y_{\alpha\beta}=R_{\alpha\gamma\beta\delta}V^\gamma V^\delta,
\end{equation}
\begin{equation}\label{37B}
X_{\alpha\beta}=^\ast R^\ast_{\alpha\gamma\beta\delta}V^\gamma
V^\delta=\frac{1}{2}\eta_{\alpha\gamma}^{\epsilon\rho}
R^\ast_{\epsilon\rho\beta\delta}V^\gamma V^\lambda,
\end{equation}
where
\begin{equation}\label{38B}
R^\ast_{\alpha\gamma\beta\delta}=\frac{1}{2}
\eta_{\epsilon\rho\gamma\delta}R_{\alpha\beta}^{\epsilon\rho}.
\end{equation}
The technique of decomposition of tensor into trace and tracefree
part along with above define tensors provide some scalars
quantities. These scalars are
$Y^{(eff)}_T,Y^{(eff)}_{TF},X^{(eff)}_T,X^{(eff)}_{TF}$ given by
\begin{equation}\label{39B}
Y_{\alpha\beta}=\frac{1}{3}Y_{T}^{(eff)}
h_{\alpha\beta}+Y_{TF}^{(eff)}\left(\chi_\alpha
\chi_\beta-\frac{1}{3}h_{\alpha\beta}\right),
\end{equation}
\begin{equation}\label{40B}
X_{\alpha\beta}=\frac{1}{3}X_{T}^{(eff)}
h_{\alpha\beta}+X_{TF}^{(eff)}\left(\chi_\alpha
\chi_\beta-\frac{1}{3}h_{\alpha\beta}\right).
\end{equation}
By using $f(R)$ field equations  and (\ref{34*B}) in Eqs.
(\ref{36B}) as well as (\ref{37B}), we get
\begin{equation}\label{41B}
Y_T^{(eff)} =4\pi(\mu^{(eff)}+3P_r^{(eff)}-2\Pi^{(eff)}),\quad
Y_{TF}^{(eff)}=\varepsilon-4\pi\Pi^{(eff)},
\end{equation}
\begin{equation}\label{42B}
X_T^{(eff)}=8\pi\mu^{(eff)},\quad
X_{TF}^{(eff)}=-\varepsilon-4\pi\Pi^{(eff)}.
\end{equation}
Equations (\ref{5B}), (\ref{6B}), (\ref{7B}), (\ref{12B}) and
(\ref{34*B}) provide
\begin{equation}\label{43B}
\frac{3m}{C^3}=4\pi(\mu^{(eff)}-\Pi^{(eff)})-\varepsilon,
\end{equation}
which combined with (\ref{31B}) and (\ref{41B}) give
\begin{equation}\label{44B}
Y_{TF}^{(eff)}=-8\pi\Pi^{(eff)}+\frac{4\pi}{C^3}\int^r_0
C^3\left(D_C \mu^{(eff)}-3q^{(eff)}\frac{U}{CE}\right)C^\prime dr.
\end{equation}

It is mentioned here that $Y^{(eff)}_{TF}$ in the above equation
differs from the $Y^{(eff)}_{TF}$ calculated for static case
\cite{6abb}. This is due to the presence of Weyl tensor, anisotropy
of effective pressure, effective energy density inhomogeneity and
the effective dissipative variables. Similarly, the scalar $X_{TF}$
is obtained as
\begin{equation}\label{45B}
X_{TF}^{(eff)}=-\frac{4\pi}{R^3}\int^r_0 C^3\left(D_C
\mu^{(eff)}-3q^{(eff)}\frac{U}{CE}\right)C^\prime dr.
\end{equation}

\subsection{Structure Scalars as Evolution Parameters}

Here, we discuss how a structure scalar can works as evolution
parameter. From Eqs.(\ref{41B})-(\ref{45B}) we evaluate the
following results
\begin{eqnarray}\label{aB}
&&X_T^{(eff)}=X_{T}^{m}+X_{T}^{D}=8\pi\mu^{(eff)}=8\pi(\mu^{m}+\mu^{D}),\\\nonumber
&&X_{TF}^{(eff)}+Y_{TF}^{(eff)}=(X_{TF}^{m}+Y_{TF}^{m})+(X_{TF}^{D}+Y_{TF}^{D})\\\label{b}
&&=-8\pi \Pi^{eff}=-8\pi(\Pi^{m}+\Pi^{D}),\\\label{cB}
&&X_{TF}^{(eff)}-Y_{TF}^{(eff)}=2\varepsilon,\\\nonumber
&&Y_{T}^{(eff)}-\frac{X_{T}^{(eff)}}{2}+X_{TF}^{(eff)}+Y_{TF}^{(eff)}
=4\pi(3P_{r}^{(eff)})\\\label{dB} &&=12\pi(P_{r}^{m}+P_{r}^{D}).
\end{eqnarray}

Equations (\ref{aB})-(\ref{dB}) describe the characteristics of
structure scalars in the dynamics of evolving cluster of stars. The
scalar $X^{(eff)}_{T}$ determines energy density due the presences
of  stars as well as non-baryonic matters. The quantities
$X^{(eff)}_{TF}$ as well as $Y^{(eff)}_{TF}$ control anisotropic
effects of pressure and tidal forces effects (Weyl tensor). The
scalar $Y^{(eff)}_{T}$ together with structure scalars
($X^{(eff)}_{T},~X^{(eff)}_{TF}$) and ($Y^{(eff)}_{TF}$) describe
the radial pressure. If the distribution within the cluster is
experiencing locally isotropic effects, it follows from Eq.(\ref{b})
that
\begin{equation}\label{eB}
X_{TF}^{(eff)}=-Y_{TF}^{(eff)}
\end{equation}
and also in addition, if the fluid is conformlly flat
($\varepsilon=0$), then Eq.(\ref{cB}) implies
\begin{equation}\label{fB}
X_{TF}^{(eff)}=Y_{TF}^{(eff)}
\end{equation}
Equations (\ref{eB}) and (\ref{fB}) show contradiction and this
implies that spherically symmetric distribution of cluster is either
isotropic or conformally flat in $f(R)$ gravity. It also predicts
that both the forces anisotropic stress and tidal force preserve
each other. Moreover, Eq. (\ref{dB}) shows that for isotropic
evolution case, the total pressure of distribution will be
calculated by $X^{(eff)}_T$ and $Y^{(eff)}_T$, whereas if the
cluster is made up of dust clouds (pressureless case),
$X^{(eff)}_T=2Y^{(eff)}_T$.

\section{Starobinsky Model}

As $f(R)$ gravity conformal with DE and DM for cluster as well as
stellar scales. Starobinsky introduced second-order curvature terms
in the gravitational action to derive Einstein equations associated
to quantum one loop distribution \cite{12a}. He established a model
for describing cosmic exponential expansion and present-time cosmic
acceleration for early time as well as present time of power-law
inflation. The model is
\begin{equation}
f(R)=R+\epsilon R^2,
\end{equation}
where $\epsilon=\frac{1}{M^2}$ and $M$ has the mass with
dimension. Later at $95\%$ confidence level, Cembrons \cite{8ab}
proved that this model solve DM problems according to WMAP data for
$\epsilon\leq\frac{1}{M^2}=2.3\times {10}^{22}Ge/V^2$ with mass
$M=2.7\times {10}^{-12}GeV$ at lower bound. By using square-
curvature terms he amend Einstein gravity at higher level of
energies (ultraviolet level). He introduced a scalar graviton as a
new scalar degree of freedom and showed that interaction between
standard model particles with scalar graviton produces thermal
abundance which indicates non-baryonic DM regions. It is also shown
that scalar degree of freedom is responsible for Yukawa force of
attraction between two non-baryonic particles of different masses if
the scalar mass $M\leq{10}^{-12}GeV$. Further, to conserve the model
upto Big Bang Nucleosynthesis temperature, the compel on mass is
$M\leq {10}^{-12}eV$.

For this model field equations turn out to be
\begin{eqnarray}\nonumber
\mu^{(eff)}&=&\left[\frac{-A^2f}{2}-\left(R^{\prime\prime}-\frac{B\dot
    B \dot R}{A^2}-\frac{2B^\prime R^\prime}{B}-B\dot B \dot
R\right)2\epsilon+KA^2\mu^{m}\right]\\\label{a*B}
&\times&\frac{1}{R(1+\epsilon
    R)}+\frac{A^2R}{2},\\\nonumber
P_r^{(eff)}&=&\left[\frac{B^2f}{2}-\left(R^{\prime\prime}-\frac{2B\dot
    B \dot R}{A^2}-\frac{2B^\prime
    R^\prime}{B}\right)2\epsilon+KB^2P^{m}_r\right]\\\label{b*B}
&\times&\frac{1}{R(1+\epsilon
    R)}-\frac{B^2 R}{2},\\\nonumber
P_\perp^{(eff)}&=&\left[\frac{R^2f}{2}+\left(\frac{R\dot
    R^2}{A^2}-\frac{RR^{\prime^2}}{B^2}-\frac{R\dot
    R^2}{A^2}\right)2\epsilon+KP^{m}_\perp R^2\right]\\\label{c*B}
&\times&\frac{1}{R(1+\epsilon
    R)}+\frac{2\epsilon R R^{\prime^2}}{B^2}-\frac{R^3}{2},\\\nonumber
q^{(eff)}&=&\left[\left(\dot R^\prime-\frac{A^\prime\dot
    R}{A}-\frac{\dot B R^\prime}{B}-\frac{2A^\prime \dot
    R}{A}-\frac{\dot B
    R^\prime}{B}\right)2\epsilon+KABq^{m}\right]\frac{1}{R(1+\epsilon R)}.\\\label{d*B}
\end{eqnarray}

\section{Evolutions of cluster of stars}

In this section, we discuss some possible simple types of evolving
clusters in the framework of $f(R)$ gravity.

\subsection{Homogenous Density and Isotropic Case}

The homogenous evolution means the evolving system shows density
homogeneity behavior whereas isotropic case is related to isotropic
pressure distribution. To describe behavior of density homogeneity
in the evolution, we firstly described a differential equation for
Weyl tensor and density \cite{8b}.  The Weyl tensor is related to
Ricci tensor by Bianchi identities
\begin{equation}
C^\eta_{\alpha\beta\kappa;\eta}=R_{\kappa[\alpha;\beta]}-\frac{1}{6}g_{{\kappa}[{\alpha}}R_{{,\beta]}}.
\end{equation}
The above relation can be converted in terms of effective
energy-momentum $T_{\mu\nu}^{(eff)}$ by using using field equations
given as
\begin{equation}\label{z*B}
C^\eta_{\alpha\beta\kappa;\eta}=T^{(eff)}_{\kappa[\alpha;\beta]}
-\frac{1}{6}g_{{\kappa}[{\alpha}}T^{(eff)}_{{,\beta]}}.
\end{equation}
Equation (\ref{33*B}) implies
\begin{equation}\nonumber
\mu^\beta C^\eta_{\alpha\beta\kappa;\eta}+\mu^\beta_{;\eta}
C^\eta_{\alpha\beta\kappa;\eta}=\theta E_{\alpha\kappa}+u^\mu
E_{\alpha\kappa;\mu}-u_{\kappa;\eta}E^\eta_\alpha-u_\kappa
E^\eta_{\alpha;\eta},
\end{equation}
which on contraction with $h^\alpha_\mu h^\kappa_\nu
u^\beta\chi^\mu\chi^\nu$ gives
\begin{eqnarray}\nonumber
h^\alpha_\mu h^\kappa_\nu \mu^\beta \chi^\mu \chi^\nu
C^\eta_{\alpha\beta\kappa;\eta}&=&\frac{4\theta}{3}E_{\mu\nu}\chi^\mu
\chi^\nu-u_{\nu;\eta}E^\eta_\mu \chi^\mu \chi^\nu+u^\beta
E_{\alpha\kappa;\beta}h^\alpha_\mu h^\kappa_\nu \chi^\mu \chi^\nu \\
\label{wB} &+& h_{\mu\nu}\sigma^{\kappa\beta}E_{\kappa\beta}\chi^\mu
\chi^\nu-\sigma_{\kappa\mu}
E^\kappa_\nu\chi^\mu\chi^\nu-\sigma^{\kappa\nu}
E^\kappa_\mu\chi^\mu\chi^\nu.
\end{eqnarray}
The effective energy-momentum tensor becomes
\begin{eqnarray}\nonumber
&&h^\alpha_\mu h^\kappa_\nu u^\beta
T^{(eff)}_{\kappa\alpha;\beta}=\left[(P_{\perp}^{(eff)})_{;\beta}u^\beta
h_{\mu\nu}  +(\Pi^{(eff)}\chi_{\kappa}
\chi_{\alpha})_{;\beta}u^\beta
h^{\kappa}_{\nu}h^{\alpha}_{\mu}\right.\\\label{zB}
&&\left.+q^{(eff)}_\mu a_\nu+q^{(eff)}_\nu a_\mu \right],\\\nonumber
&&h^\alpha_\mu h^\kappa_\nu u^\beta T_{\kappa
\beta;\alpha}^{(eff)}=\left[(-\mu^{(eff)}+3P_{\perp}^{(eff)})\left(\sigma_{\mu\nu}+\frac{\Theta
h_{\mu\nu}}{3}\right)-q^{(eff)}_{;\alpha}h^\alpha_\mu \chi_\nu\right.\\\label{yB}
&&\left.+\Pi^{(eff)}_{;\beta}\mu^\beta\chi_\kappa\chi_{\beta;\alpha}u^\beta
h^\alpha_\mu h^\kappa_\nu\right],\\\label{xB} &&h^\alpha_\mu
h^\kappa_\nu u^\beta g_{[\alpha}
T^{(eff)}_{;\beta]}=h_{\mu\nu}[-\mu^{(eff)}+3P_{\perp}^{(eff)}+\Pi^{(eff)}]_{;\beta}u^\beta.
\end{eqnarray}
From the Eqs. (\ref{z*B}) and (\ref{zB})-(\ref{xB}), we get
\begin{eqnarray}\nonumber
&&\left[\left(\varepsilon+\frac{4\pi\mu^{(eff)}}{2}-\frac{4\pi\Pi^{(eff)}}{2}-\right)_{,\mu}u^\mu
+\frac{12\pi
q^{(eff)}C'}{2BC}+\left(\varepsilon+\frac{4\pi\mu^{(eff)}}{2}\right.\right.\\\label{pB}
&&\left.\left. -\frac{4\pi\Pi^{(eff)}}{2}+4\pi
P_{r}^{(eff)}\right)(\Theta+\sigma)\right]=0.
\end{eqnarray}
Similarly, contraction of Weyl tensor and Ricci tensor with
$u^\kappa u^\beta h^\alpha_\mu \chi^\mu$ yields
\begin{eqnarray}\nonumber
&&\left(\varepsilon+\frac{4\pi\mu^{(eff)}}{2}-\frac{4\pi\Pi^{(eff)}}{2}\right)_{,\mu}\chi^{\mu}
-\frac{3C'}{BC}\left(\frac{4\pi\Pi^{(eff)}}{2}-\varepsilon\right)\\\label{qB}
&&-4\pi q^{(eff)}(\sigma+\Theta)=0.
\end{eqnarray}
The above two Eqs. (\ref{pB}) and (\ref{qB}) explain the relations
for density inhomogeneity, Weyl tensor, anisotropy and dissipation
under the effects of gravitation effects related to exotic matter.

Equations (\ref{42B}) and (\ref{qB}) give
\begin{equation}\label{rB}
\left(X^{(eff)}_{TF}+4\pi\mu^{(eff)}\right)'=-X_{TF}^{(eff)}\frac{3C}{C}
+4\pi q^{(eff)} B(\theta-\sigma).
\end{equation}
The derived relation describes the density homogeneity through
spatial derivative of effective density contribution. If
$(\mu^{(eff)})'=(\mu^{(m)}+\mu^{D})'=0$, the density distribution is
homogenous in the cluster. This implies that the homogenous
evolution of cluster totally depends upon the behavior of baryonic
as well as  non-baryonic density.

Now we discuss this phenomenon in more details and find factors as
well as condition for such type of evolution. Firstly, we consider non
dissipative case $(q^{(eff)})=0$, in this situation Eq.(\ref{rB})
implies
\begin{equation}\label{47B}
X_{TF}^{(eff)}=0\Rightarrow \mu^{{\prime}{(eff)}}=0.
\end{equation}
This shows that, in non dissipative case,  homogenous evolution of
the cluster is controlled by scalar $X_{TF}^{(eff)}$. This condition
along with (\ref{42B}) implies
\begin{equation}
\varepsilon+\Pi^{(eff)}=0,
\end{equation}
which is possible if either $\varepsilon=4\pi\Pi^{(eff)}=0$, or
$\varepsilon=-4\pi\Pi^{(eff)}$. The first case indicates that the
evolution is also isotropic as well as conformally flat while the
second one shows that effective anisotropic effects and tidal forces
effects (describe by $\varepsilon$) are balancing each other. The first
case is not possible as we have shown in the section $4.1$ that the
fluid is either isotropic or conformally flat.

If the evolving cluster of stars create dissipation then we have
\begin{equation}\label{48B}
X_{TF}^{(eff)}=0\Rightarrow
\mu^{{\prime}{(eff)}}=q^{(eff)}B(\theta-\sigma)=q^{(eff)}B\frac{3\dot
R}{R}.
\end{equation}
This shows that homogenous evolution is controlled by effective
dissipative along with scalar $X_{TF}^{(eff)}=0$, expansion
parameter and shear tensor. For expansion as well as shear free
evolution the dissipation of the system will not played any role for
evolution and the evolution has homogenous density along with
$X_{TF}^{(eff)}=0$.

From the above discussion, it can be noticed that non dissipative
homogenous evolution of cluster of stars containing non-baryonic
matter (DM) depends upon two conditions, that is  isotropic
and conformally flat or anisotropic pressure balancing tidal forces.
The dissipative homologous evolution depends upon dissipative
factors due to matter as well as DM along with expansion
and shear effects. Here the dissipation due to DM might be
some sort of squandering of non-baryonic particles. The dissipative
cluster having expansion and shear effects shows density
inhomogeneity during evolution.

\subsection{The Homologous Evolution}

The homologous evolution means the evolving cluster of stars is
preserving position, structure or origin. To discuss this case, we
explore the behavior of evolving velocity $U$. From Eqs. (\ref{d*B})
and (\ref{26*B}), we have
\begin{equation}\label{54B}
D_C\left(\frac{U}{C}\right)=-\frac{4\pi}{RE(1+\epsilon
R}\left[\left(\dot R^\prime-\frac{3A^\prime \dot R}{A}-\frac{2\dot
BR^\prime}{B}\right)2\epsilon+KABq^{(m)}\right]+\frac{\sigma}{C},
\end{equation}
after integrating, we get
\begin{eqnarray}\nonumber
&&U=\tilde{a}(t)C+\int_0^r\left(\frac{\sigma}{C}-\frac{4\pi}{RE(1+\epsilon
R)}\left[KABq^{m}+\left(\dot R^\prime-\frac{3A^\prime\dot R}{A}
\right.\right.\right.\\\label{55} &&\left.\left.\left.-\frac{2\dot B
R^\prime}{B}\right)2\epsilon\right]\right)C^\prime dr.
\end{eqnarray}
Here $\tilde{a}(t)$ is an integrating function and we have
\begin{eqnarray}\nonumber
&&U=\frac{U_{\Sigma}}{C_{\Sigma}}C-\int_0^r\left(\frac{\sigma}{C}-\frac{4\pi}{RE(1+\epsilon
R)}\left[KABq^{m}+\left(\dot R^\prime-\frac{3A^\prime\dot R}{A}
\right.\right.\right.\\\label{56aB} &&\left.\left.\left.-\frac{2\dot
B R^\prime}{B}\right)2\epsilon\right]\right)C^\prime dr.
\end{eqnarray}
The above equation governed the homologous evolution of the cluster.
If the integral of (\ref{56aB}) vanishes, we have
\begin{equation}\label{57B}
U=\tilde{a}(t)C,\quad \tilde{a}(t)\equiv\frac{U_\Sigma}{C_\Sigma},
\end{equation}
or
\begin{equation}\label{58aB}
U\sim C
\end{equation}
which is a feature of homologous evolution in Newtonian
hydrodynamics \cite{9}. It can be observed from (\ref{56aB}), if the cluster
does not show shear as well as dissipative effects, then the
velocity of evolving cluster is monitored by high curvature terms
(DM terms) in the radial direction. Thus, there are two
possibilities for homologous condition (\ref{57B})
in non dissipative situation, for shear-free case $(\sigma=0)$,
the effects of high curvature terms vanishes for
$f(R)\sim R$. Whereas in shear case the DM
terms canceled out the shear effects and the integral in
Eq.(\ref{56aB}) vanishes. It can be noticed that, in GR, such type of
evolution take place in shear-free and non-dissipative case \cite{6abb} but in
higher curvature scenario the homologous evolution can take place in
shear case as well.

The relativistic homologous condition is derived by considering two
concentric shells having areal radius $C_1$ and $C_{2}$ at
$r=r_{1}=constant$, and $r=r_{2}=constant$, respectively, given by
\cite{6abb}
\begin{equation}\label{58aaB}
\frac{C_1}{C_{2}}=constant.
\end{equation}
The critical factor that can be noticed here is that condition
(\ref{58aB}) does not implies condition (\ref{58aaB}). By applying
Eq.(\ref{58aB}) for two shells of stellar fluid $1,~2$, the
homologous condition turns out to be
\begin{equation}\label{59B}
\frac{U_{1}}{U_{2}}=\frac{A_{1}\mathcal{C}_{1}}{A_{2}\mathcal{C}_{2}}=
\frac{C_{1}}{C_{2}},
\end{equation}
this implies (\ref{58aaB}) only if $A=A(t)$. By using simple
coordinate transformation, we can convert $A=constant$ which is a
property of geodesic fluid $(a=\frac{A'}{AB}=0)$. This shows that
for non-relativistic system the condition (\ref{58aaB}) satisfy
whenever $U\sim C$, whereas for relativistic case the condition
(\ref{58aB}) implies (\ref{58aaB}), only if the evolving system is
geodesic. Thus, from Eq.(\ref{56aB}) the quasi-homologous evolution
take place if
\begin{equation}\label{60B}
\frac{-4\pi B}{RC^\prime(1+\epsilon R)}\left[\left(\dot
R^\prime-\frac{3A^\prime \dot R}{A}-\frac{2\dot B
R^\prime}{B}\right)2\epsilon+KABq^{(m)}\right]+\frac{\sigma}{C}=0.
\end{equation}
This implies that in the absences of shear and dissipation the high
curvature terms associated to DM controlled the homologous
evolution. In both cases dissipative or non-dissipative DM
terms play a important role in controlling homologous condition.

It can be observed that according to the composition to cluster of
stars, we cannot remove dark matter effects in the evolution of
cluster that is
\begin{equation}\label{61aB}
\frac{-4\pi B}{RC^\prime(1+\epsilon R)}\left[\left(\dot
R^\prime-\frac{3A^\prime \dot R}{A}+\frac{2\dot B
R^\prime}{B}\right)2\epsilon\right]\neq0.
\end{equation}

So the stability of homologous condition occurs whenever
\begin{equation}\label{62aB}
\frac{-4\pi B}{RC^\prime(1+\epsilon R)}\left[\left(\dot
R^\prime-\frac{3A^\prime \dot R}{A}-\frac{2\dot B
R^\prime}{B}\right)2\epsilon+KABq^{(m)}\right]=-\frac{\sigma}{C}.
\end{equation}
In contrast to GR results, Eqs.(\ref{61aB}) and (\ref{62aB}) show
that the homologous evolution of cluster of star can be dissipative
and shear-free or can be non-dissipative with shear effects.

\subsection{The homogeneous expansion}

Here we describe evolution of cluster of stars with homogenous
expansion. In homogenous expansion, the rate of change of expansion
with respect to radius of cluster becomes zero. Thus (\ref{26*B})
implies
\begin{equation}\label{60*B}
4\pi
q^{(eff)}=-\frac{C^\prime}{B}\left[\frac{1}{3}D_C(\sigma)
+\frac{\sigma}{C}\right].
\end{equation}
This shows the total effective dissipation is responsible for shear
effects of homogenously expanding cluster. Since $q^{eff}=q^m+q^D$,
so in the absences of dissipation
due to matter the shear effects are totally depends upon
dissipation due to DM. Now, if we apply shear free conditions, we get
\begin{equation}
4\pi q^{(eff)}=\frac{-4\pi B}{RC^\prime(1+\epsilon
R)}\left[\left(\dot R^\prime-\frac{3A^\prime \dot R}{A}-\frac{2\dot
B R^\prime}{B}\right)2\epsilon+KABq^{(m)}\right]=0.
\end{equation}
This condition along with (\ref{61aB}) imply that the total
effective dissipation of the system can be zero, in a way, if the
dissipation of DM becomes equal to other dissipation
related to baryonic matter distribution.

\subsection{Criteria of Geodesic Evolution}

During geodesic evolution the fluid particles moves with constant
velocity ($a=0$) along geodesics. So in geodesic evolution
of cluster, stars show movement with constant velocity.
Here we discuss criteria of such
type of evolution of cluster of stars \cite{6abb}. For this, we considered
simplest evolution scenario, the quasi-homologous condition
(\ref{60B}) which implies
\begin{equation}\label{62B}
\frac{-4\pi B}{R(1+\epsilon R)}\left[\left(\dot
R^\prime-\frac{3A^\prime \dot R}{A}-\frac{2\dot B
R^\prime}{B}\right)2\epsilon+KABq^{(eff)}\right]=\frac{\sigma
C^\prime}{C},
\end{equation}
putting this expression in (\ref{26*B}) gives
\begin{equation}\label{63B}
(\Theta-\sigma)^\prime = 0.
\end{equation}
Then by using Eqs. (\ref{9B}) and (\ref{11B}) we get
\begin{equation}\label{64*B}
(\Theta-\sigma)^\prime =  \left(\frac{3}{A}\frac{\dot
C}{C}\right)^\prime=0.
\end{equation}
Equation (\ref{59B}) implies $C$ is a separable function that is
$C=C_{I}(t)C_{II}(r)$ which along with (\ref{64*B}) gives
\begin{equation}\label{65B}
A^\prime =0.
\end{equation}
By the re-parametrization of the coordinate $r$, (\ref{65B}) yield
$A=1$ which shows geodesic fluid.

In the inverse situation the geodesic fluid also satisfies
homologous condition. For $A=1$ we get
\begin{equation}\label{66B}
\Theta-\sigma=\frac{3\dot C}{C},
\end{equation}
which implies ($\theta-\sigma)^\prime=0$, near to center ($C \sim
r$). Taking r-derivatives of (\ref{66B}) successively we obtain, at
($C \sim r$),
\begin{equation}\label{67B}
\frac{\partial^n(\Theta-\sigma)}{\partial r^n}=0.
\end{equation}
Here we consider $(\Theta-\sigma)^\prime$ is of class $M^p$, i.e, by
Taylor series expansion around the center, the fluid becomes
homologous (\ref{63B}) when we take zero value analytically at the
center. Thus geodesic and homologous condition imply each other .
\section{Physical Significance of Starobinsky Model}

\textbf{In order to study the physical significance of Starobinsky model ($f(R)=R+\epsilon R^2$)
for self-gravitating structures, we explore behavior of parameter $\epsilon$ for a compact self-gravitating stellar. For this, we use simple model base upon spatial astral density (an idea similar to de Vaucoulour's account in the exterior region
avoiding definite center). Jaffe \cite{1g} and Hernquist \cite{2g} were the first to introduce such type of two models having central astral densities proportional to $r^{-2}$ and $r^{-1}$. These models can provide a family of density profiles
associated with diverse central slopes given as
\begin{equation}\label{AA}
\mu(r)=\frac{(3-\lambda)M\alpha}{4\pi r^{\lambda}(\alpha+r)^{4-\lambda}},
\end{equation}
where $M$ shows the whole mass and $\alpha$ is a scaling radius. The mass distribution is proportional to $r^{\lambda}$ at the center. The value of $\lambda$ is limited to the interval $[0,3)$ and for $\lambda=2,~ \lambda=1,$ the model reduces to Jaffe and Hernquist models, respectively.
In the present study, we use Hernquist model with $\lambda=1$.
We apply Krori-Barua anstaz \cite{3g}
technique to the metric functions associated to interior spacetime. That is, we chose interior metric functions as
\begin{equation}\label{BB}
A=e^{a},\quad B=e^{b},\quad C=r,
\end{equation}
with
$a=\tilde{B}r^2+\tilde{C}$ and
$b=\tilde{A}r^2$.}

\textbf{To connect realistic compact stellar model with static
and asymptotically flat exterior region, we consider the Schwarzschild metric as exterior geometry given by}
\begin{equation}\label{CC}
d s^{2}=\left(1-\frac{2 M}{r}\right) d t^{2}-\left(1-\frac{2 M}{r}\right)^{-1} d r^{2}
-r^{2}\left(d \theta^{2}+\sin ^{2} \theta d \phi^{2}\right).
\end{equation}
\textbf{Now we applying matching condition upon interior and exterior spacetimes.
The continuity of interior and exterior line elements at the condition $r=R$ gives}
\begin{equation}\label{DD}
g_{t t}^{-}=g_{t t}^{+}, \quad g_{r r}^{-}=g_{r r}^{+},
\quad \frac{\partial g_{t t}^{-}}{\partial r}=\frac{\partial g_{t t}^{+}}{\partial r},
\end{equation}
\textbf{where the + and - signs denote the star's exterior and interior surfaces,
respectively. Thus, we obtain from Eq.(\ref{DD})}
\begin{eqnarray}\label{EE}
&&\tilde{A}=-ln\left(1-\frac{2 M}{r}\right)\frac{1}{R^2},\quad
\tilde{B}=\left(1-\frac{2 M}{r}\right)^{-1}\frac{M}{R^3},\\\label{GG}
&&\tilde{C}=ln\left(1-\frac{2 M}{r}\right)-\left(1-\frac{2 M}{r}\right)^{-1}\frac{M}{R}.
\end{eqnarray}
\textbf{By using mass-radius analysis of various compact stellar,
we can connect the values of interior metric functions (\ref{BB})
with the observational data \cite{1f}.
Zhang.et.al. \cite{2f} discussed globular cluster binary
source $4U 1820-30$ for neutron star. They evaluated testified
mass of the direction as $M\simeq2.2M\odot$. Guver et al. \cite{3f}
studied $4U 1820-30$  and measure mass and radius of neutron
star with $1\sigma$ error as $R=9.11\pm0.4$km and $M=1.58\pm0.06M\odot$.
However, as compare to Zhang.et.al. results, an upper bound limit in
this calculation is invariant. Because there is a firm confusion in size
of mass and radius of dense star. Thus, the calculated values of
$\tilde{A}, \tilde{B}$ and $\tilde{C}$ for $4U 1820-30$
having mass $M\simeq2.2M\odot$ and radius $R\simeq10.0$ km are given by}
\begin{eqnarray}\label{II}
&&\tilde{A}= 0.010906441192 km^{-2},\\\label{GG}
&&\tilde{B}=0.0098809523811 km^{-2},\\\label{KK}
&&\tilde{C}=-2.0787393571141.
\end{eqnarray}
\textbf{By using the values of $\tilde{A},~\tilde{B}$ and $\tilde{C}$
in field equations and Eqs.(\ref{a*B})-(\ref{d*B}),
we obtain values of density, radial as well as tangential pressures
of dark matter and baryonic matter associated to Starobinsky model. The
Fig \textbf{1.} shows the behavior of DM density in the cluster.
It can be seen that for the increasing values of $\epsilon$, the cluster
becomes more dense toward the center. Figures \textbf{2.} and \textbf{3.} describe
the radial and tangential pressures associated to DM. The radial pressure increases
while tangential pressure decreases toward the center for increasing values of
$\epsilon$.  The Fig \textbf{4.} shows the behavior of matter density $\mu^{m}$.
It can be observed that for increasing values of $\epsilon$ the matter density is almost
negligible in the cluster. Moreover, it can be noticed form figures \textbf{1.} and \textbf{4.}
that for increasing values of $\epsilon$, DM density increases whereas
baryonic matter density became negligible toward center.
This result is in accord to observational data,
since in the cluster or galaxies the ratio of baryonic matter
is very less as compare to DM. Figures \textbf{5.} and \textbf{6.} shows that
the radial pressure of matter increases while matter tangential pressure
decreases to vanish toward the center
for increasing values of $\epsilon$.
Hence it can be observed from the above analysis that as the value of $\epsilon$ increases
the energy density of DM
becomes dominant as compare to matter density.
The radial pressures of both matters remain increasing
while the tangential pressure decreases to zero.}
\begin{figure}[htbp]
\begin{center}
\includegraphics[scale=.45]{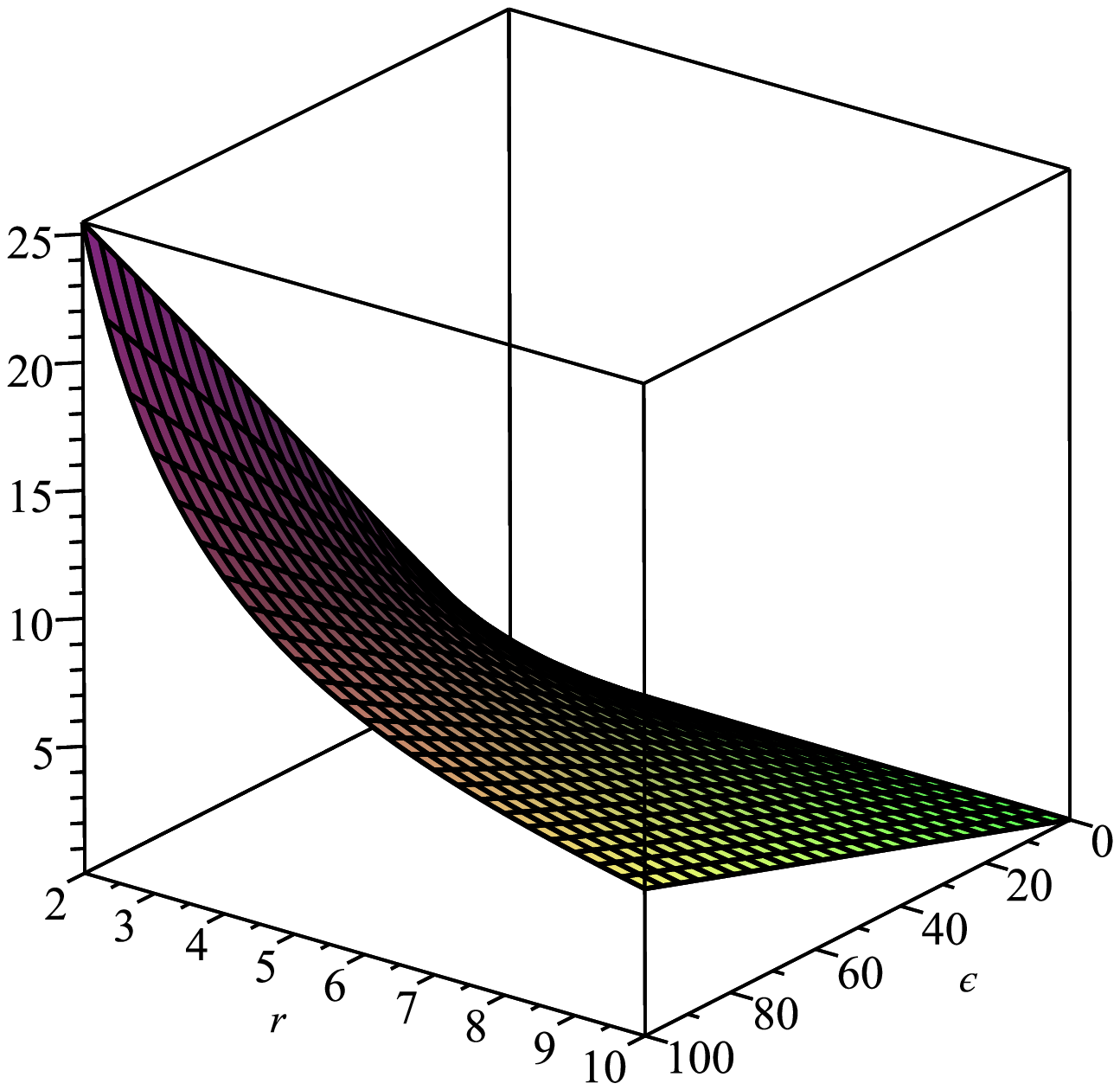}
\caption{Variation of $T^{D}_{00}=\mu^{D}$ as function of $r$ and $\epsilon$}
\label{fig 1}\quad
\includegraphics[scale=.45]{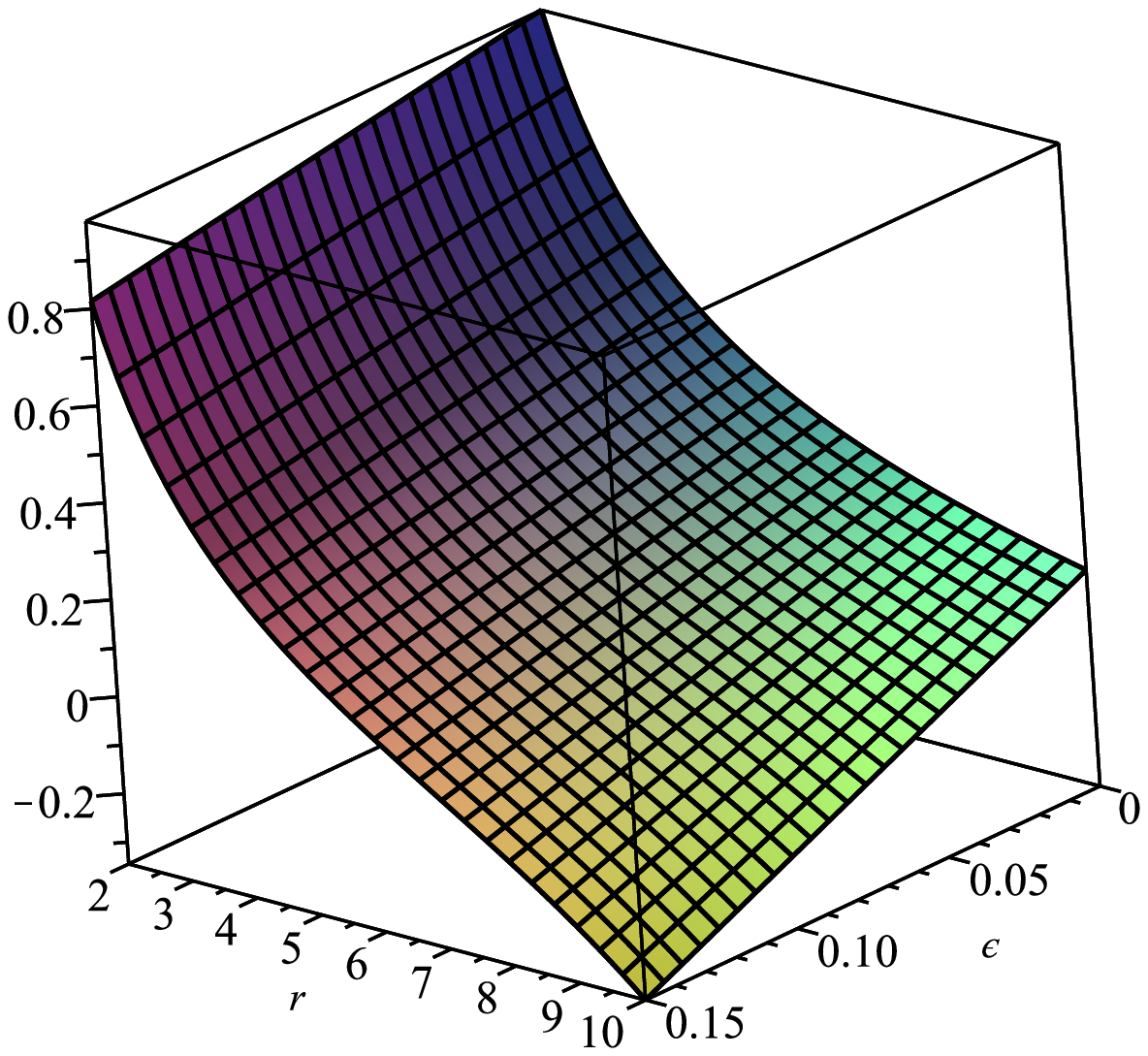}
\caption{Variation of $T^{D}_{11}=P^{D}_{r}$ as function of $r$ and $\epsilon$}
\label{fig 2}
\end{center}
\end{figure}

\begin{figure}[htbp]
\begin{center}
\includegraphics[scale=.4]{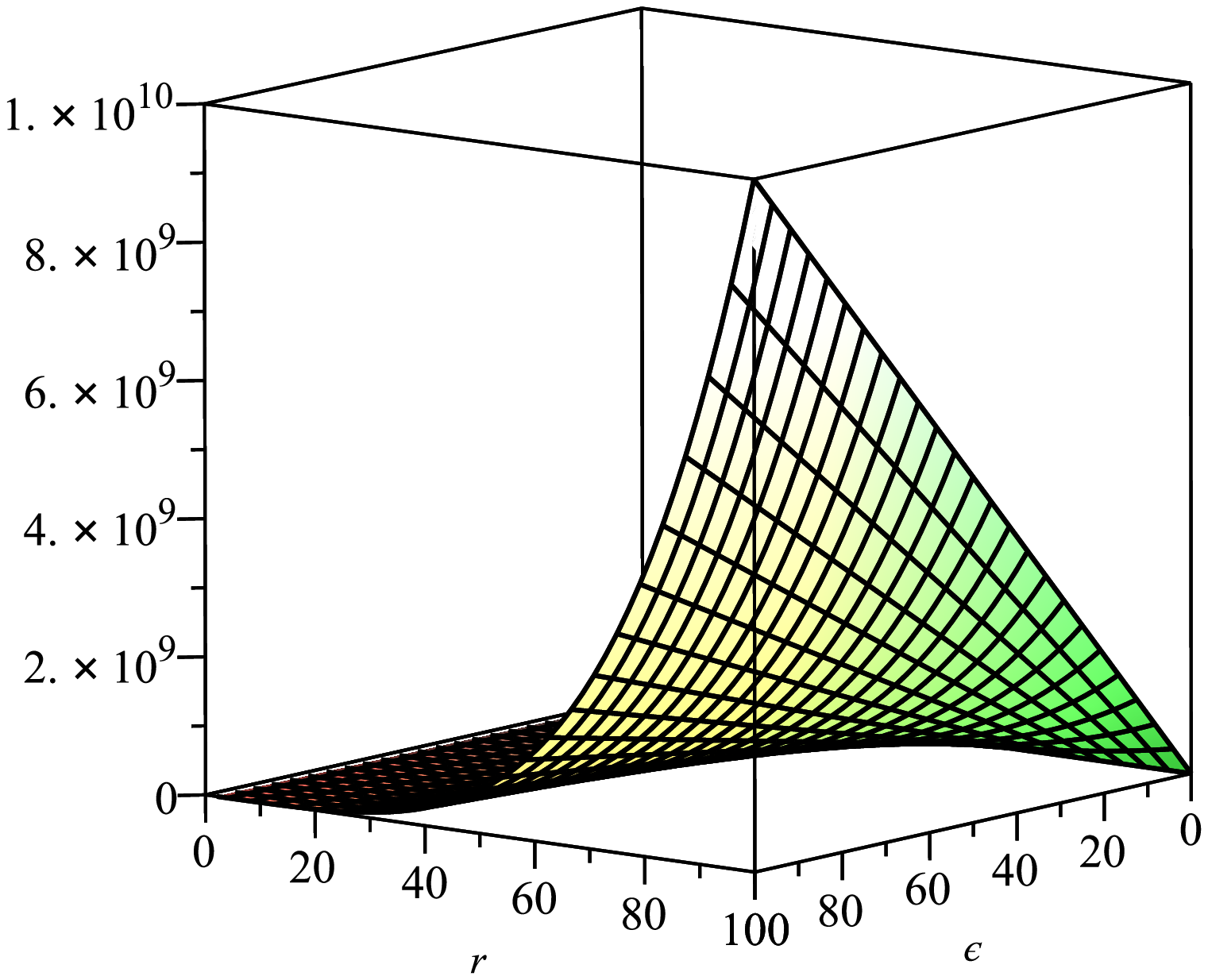}
\caption{Variation of $T^{D}_{22}=P^{D}_{\perp}$ as function of $r$ and $\epsilon$}
\label{fig 3}
\end{center}
\end{figure}
\begin{figure}[htbp]
\begin{center}
\includegraphics[scale=.5]{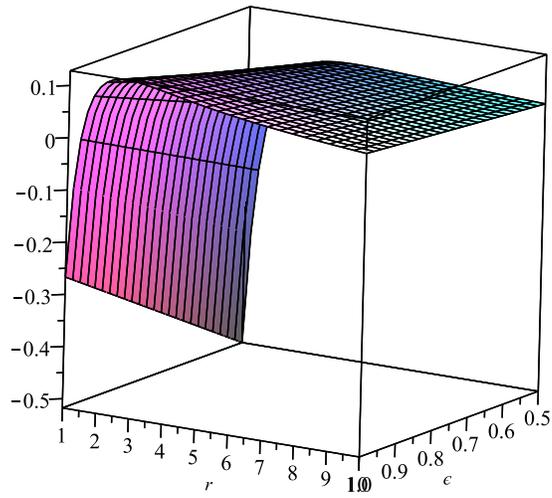}
\caption{Variation of $\mu^{m}$ as function of $r$ and $\epsilon$}
\label{fig 4}
\end{center}
\end{figure}
\begin{figure}[htbp]
\begin{center}
\includegraphics[scale=.6]{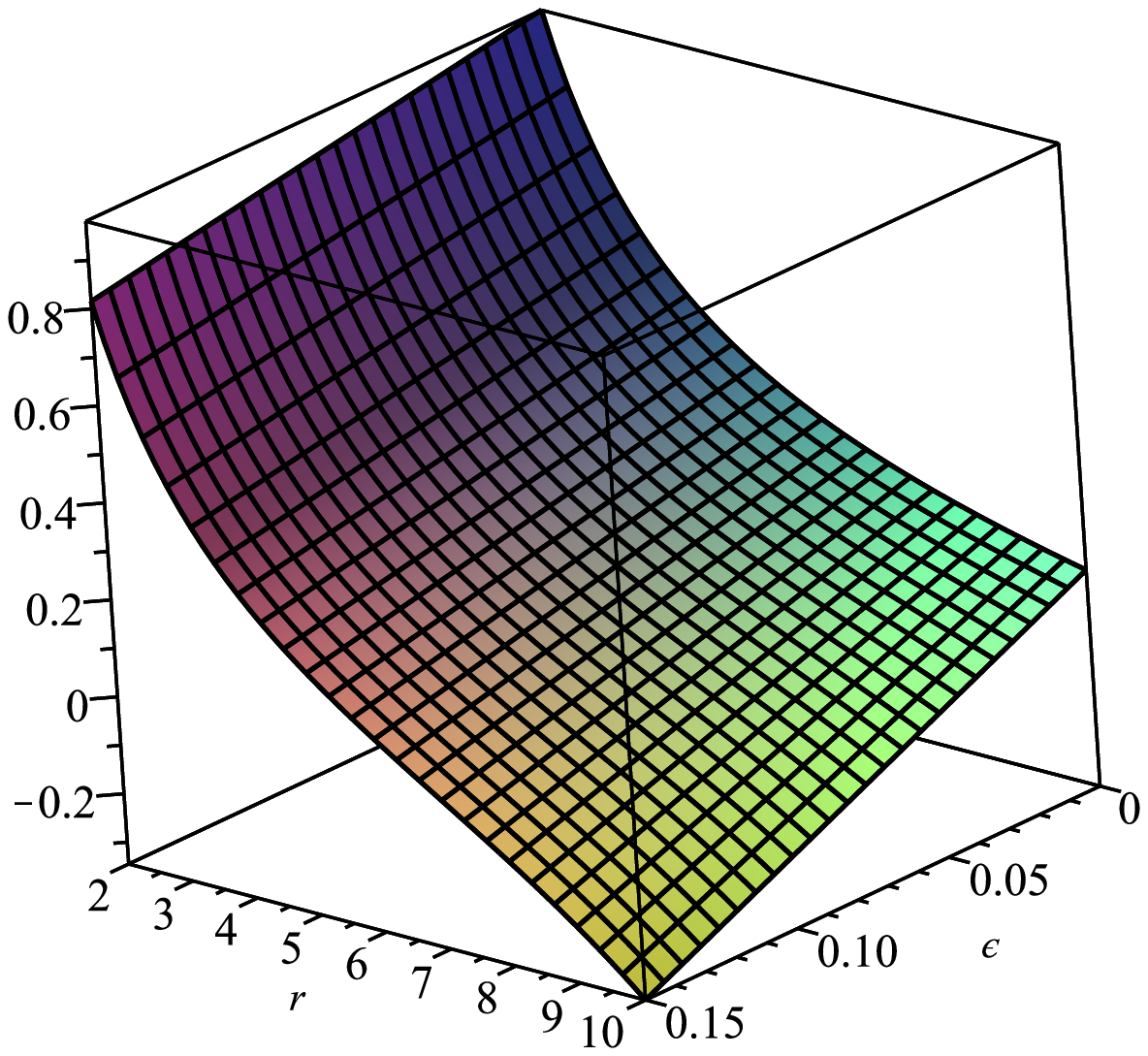}
\caption{Variation of $P^{m}_{r}$ as function of $r$ and $\epsilon$}
\label{fig 5}
\end{center}
\end{figure}
\begin{figure}[htbp]
\begin{center}
\includegraphics[scale=.5]{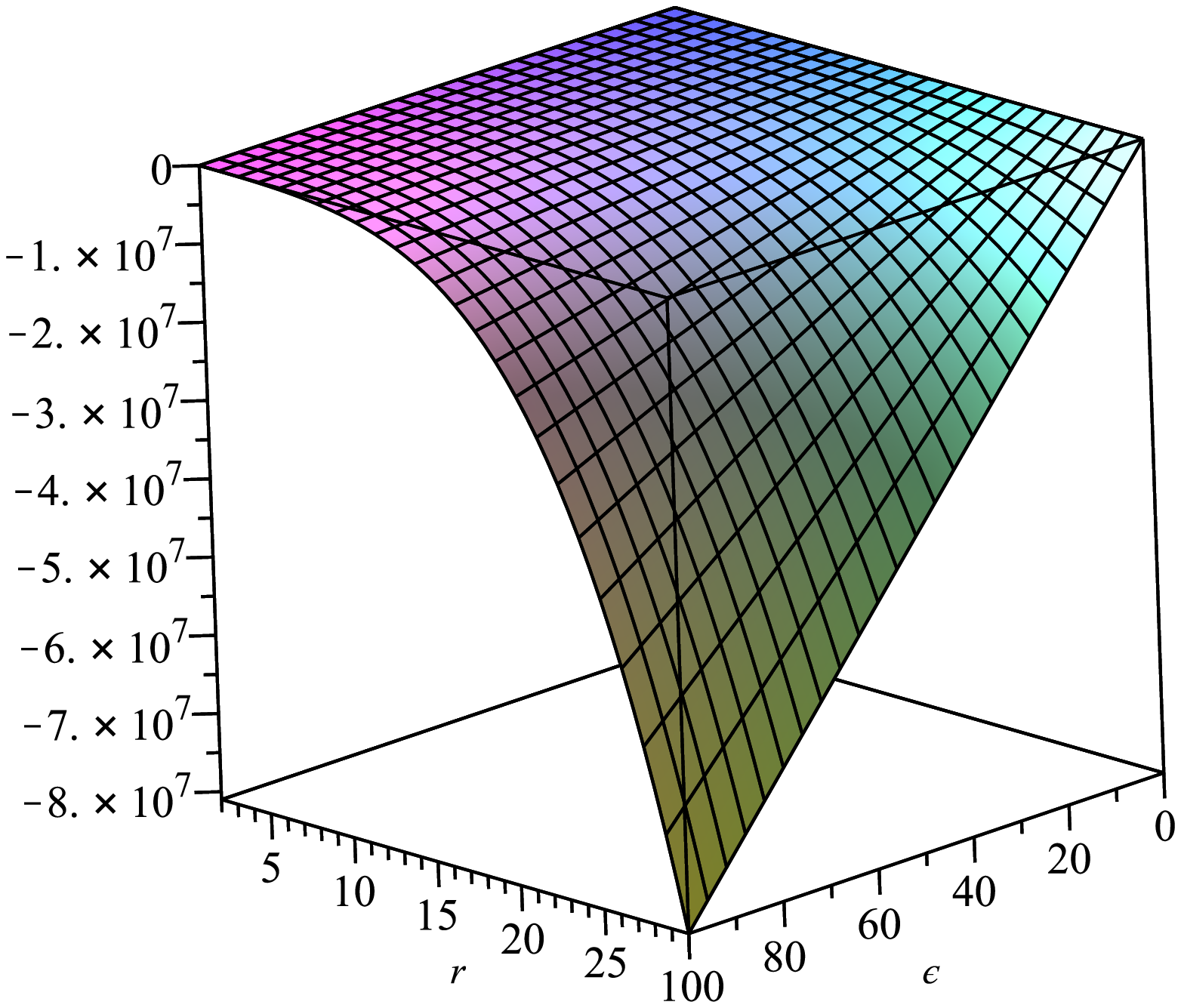}
\caption{Variation of $P^{m}_{\perp}$ as function of $r$ and $\epsilon$}
\label{fig 6}
\end{center}
\end{figure}

\section{Conclusion}

In this paper, we have studied cluster of stars in spherically
symmetrical form where the fluid particles are a combination of
baryonic (stars of clusters) and non-baryonic (DM) types matter. We have
used the concept of high curvature gravity ($f(R)$ gravity) to
include DM in the discussion. We have explored various
simple modes of evolving cluster with the help of structure scalars
in Starobinsky model, $f(R)=R+\varepsilon R^2$, to find the effects
of non-baryonic matters on the evolution of stars cluster.

The structure scalars are used for exploring dynamics of
self-gravitating structures. These factors refined the
stellar evolution mechanism by reducing the system's complexity. We
have used the generalized definition of structure scalars in $f(R)$
gravity to propose the dynamics of refine (complex less) evolution
of clusters. It has found that these scalars controls effects of
tidal forces (Weyl tensor), effective anisotropic pressure and effective
dissipative as well as inhomogeneous energy density factors.
It has also been shown that the structure scalars suggested that
structure of spherically symmetric cluster can be either isotropic or
conformally flat and the two forces anisotropic
stress and tidal force support each other.

Galactic observations indicated that the life and evolution of
clusters depend on a considerable amount of DM. This
shows that the DM can affect the evolving structure of clusters.
Firstly, we have studied evolution with homogenous
density and isotropic case. It is noticed that non-dissipative
evolution shows density homogeneity if the
anisotropic pressure of the fluid counteract with
the tidal forces effects. The dissipative case depends upon
dissipative factors due to baryonic matter as well as DM along with
expansion and shear effects. The dissipating cluster having
expansion and shear effects shows density inhomogeneity. For
expansion and shear free situation, the dissipative evolution shows
isotropic behavior with homogenous density.

We then have discussed homologous evolution for stars cluster in the
presence of non-baryonic matter. For non-relativistic  homologous
evolution, it has found that if the cluster does not show shear
as well as dissipative effects, then the velocity of evolving
cluster is monitored by high curvature terms (DM terms) in
the radial direction. So vanishes of higher curvature terms
($f(R)\sim R$) causes homologous evolution. Whereas in shear and
non-dissipative case, this type of evolution take place if the
DM terms canceled out the shear effects. It is noticed that, in
GR, such type of evolution take place in shear-free and
non-dissipative case but in higher curvature
scenario the homologous
evolution can take place in shear case as well.

The relativistic type of homologous evolution has also been
explored. It is found that in either situations dissipative or
non-dissipative DM terms play a important role in
controlling homologous condition. Moreover, in contrast to GR
results, it is concluded that the homologous evolution of cluster of
star can be dissipative and shear-free or can be non-dissipative
with shear effects.

We have explored condition for expanding cluster. It is concluded
that expanding cluster depends upon shear effects along
with dissipation due to matter and DM. The geodesic
evolution also have investigated and it is obtained
that the stars cluster moves with constant velocity. But around the
center the fluid becomes homologous which shows geodesic and
homologous imply each other.

\textbf{Finally we explore behavior of Starobinsky model for a self-gravitating stellar
$4U 1820-30$ for values of parameter $\epsilon$ toward center. It has concluded that
for large values of $\epsilon$ the density of
DM becomes dominant as compare to matter density.
The radial pressures remain dominant
and increasing for both DM and baryonic matter.
While the tangential pressures of both matter decreases to zero.
}

\end{document}